\documentclass[conference]{IEEEtran}
\usepackage{cite}
\usepackage{amsmath,amssymb,amsfonts}
\usepackage{algorithmic}
\usepackage{algorithm}
\usepackage{graphicx}
\usepackage{textcomp}
\usepackage{xcolor}
\renewcommand{\baselinestretch}{0.94}
\usepackage{booktabs} 
\usepackage{epsfig}
\usepackage[TABBOTCAP]{subfigure}
\usepackage{tabularx}
\usepackage{graphicx}
\usepackage{color}
\usepackage{xspace}
\usepackage{thumbpdf}
\usepackage{listings}
\usepackage{verbatim}
\usepackage{booktabs}
\usepackage{colortbl}
\usepackage{url}
\usepackage{nth}
\usepackage{multirow}
\usepackage{multicol}
\usepackage[english]{babel}
\usepackage[utf8]{inputenc}
\usepackage{algorithm}
\usepackage{mathtools}

\def\BibTeX{{\rm B\kern-.05em{\sc i\kern-.025em b}\kern-.08em
    T\kern-.1667em\lower.7ex\hbox{E}\kern-.125emX}}
\begin{document}
\title{Adversarial Machine Learning Attack on Modulation Classification}

\author{\IEEEauthorblockN{Muhammad Usama\IEEEauthorrefmark{1},
Muhammad Asim\IEEEauthorrefmark{1},
Junaid Qadir\IEEEauthorrefmark{1},
Ala Al-Fuqaha\IEEEauthorrefmark{2},
Muhammad Ali Imran\IEEEauthorrefmark{3}
}
\IEEEauthorblockA{\IEEEauthorrefmark{1}Information Technology University, Lahore, Punjab, Pakistan.}
\IEEEauthorblockA{\IEEEauthorrefmark{2}Hamad Bin Khalifa University, Qatar}
\IEEEauthorblockA{\IEEEauthorrefmark{3}University of Glasgow, Scotland, UK}

Email: \IEEEauthorrefmark{1}(muhammad.usama, msee16001, junaid.qadir)@itu.edu.pk, \IEEEauthorrefmark{2}aalfuqaha@hbku.edu.qa,
\IEEEauthorrefmark{3}muhammad.imran@glasgow.ac.uk
}

\maketitle

\begin{abstract}
Modulation classification is an important component of cognitive self-driving networks. Recently many ML-based modulation classification methods have been proposed. We have evaluated the robustness of 9 ML-based modulation classifiers against the powerful Carlini \& Wagner (C-W) attack and showed that the current ML-based modulation classifiers do not provide any deterrence against adversarial ML examples. To the best of our knowledge, we are the first to report the results of the application of the C-W attack for creating adversarial examples against various ML models for modulation classification.

\end{abstract}


\section{Introduction}

The success of machine learning (ML) in computer vision and speech processing has motivated the networking community to consider deploying ML for the automation of networking operations. Recently, new networking paradigms like cognitive self-driving networks \cite{feamster2017and} and most recently knowledge defined networking \cite{mestres2017knowledge} have also emerged that depend on and facilitate the extensive utilization of ML schemes for conducting networking tasks. Recently ML has successfully applied on different cognitive self-driving networking tasks such as modulation classification \cite{o2016convolutional} and representation learning of radio signals \cite{o2016unsupervised}.

Although, ML schemes especially deep neural networks (DNN) have outperformed traditional networking schemes in many networking tasks, it has been shown recently that DNN and other ML schemes lacks robustness against ``\textit{adversarial examples}'', which are defined as inputs to the ML model specially crafted by an adversary to cause a malfunction in the performance of the ML model. These adversarial examples are generated by adding small typically-imperceptible perturbations to the legitimate examples for the express purpose of misleading the ML model towards the production of wrong results and to increase the prediction error of the model. 

Based on the adversary's knowledge, adversarial attacks are classified into two major categories: white-box attacks and black-box attacks. In \textit{white-box attacks}, it is assumed that the adversary has perfect knowledge about the victim model, whereas in \textit{black-box attacks} it is assumed that adversary has no information about the victim model and the adversary can only query the deployed ML model for a response and to later use this information for crafting adversarial examples.

More formally, an adversarial example $x^*$ is crafted by adding a small imperceptible perturbation $\delta$ to the test example $x$ of the deployed trained classifier $f(.)$. The perturbation $\delta$ is computed by approximating iteratively the nonlinear optimization problem given in equation 1 until the crafted adversarial example gets classified by the trained ML classifier $f(.)$ in a wrong class $t$. 

\begin{equation}\label{eq1}
    x^* = x + \arg \underset{\eta{_x}}{\text{min}} \{\|\eta\|: f(x + \eta) = t\} 
\end{equation}

Adversarial examples are a direct  consequence of an unsafe assumption in ML that distribution encountered by the ML model in training phase will also be encountered in the test phase of the ML model.

The effects of adversarial ML examples in cognitive self-driving networks have not been explored properly in the literature. In this paper, we have performed an adversarial attack on ML classifiers performing the task of modulation classification, which is an important application in cognitive self-driving networks. Our results clearly highlight that a small optimally-calculated adversarial perturbation for the test example can cause a serious drop in performance of the classification output of the ML model. This paper also highlights the vulnerability and brittleness associated with the ML models used in the cognitive self-driving networks. 

The \textbf{major contributions} of this work are: 

\begin{itemize}
    
    \item We have performed an adversarial ML attack on 9 ML-based modulation classifiers to highlight the vulnerability of these modulation classifiers to adversarial perturbation;
    
    \item We demonstrate the \textit{transferability} phenomenon in the setting of modulation classifiers by showing that an adversarial example compromising one ML scheme will also be to evade other ML schemes with high probability;
    
    \item \textit{To this best of our knowledge, this is the first experiment where the Carlini \& Wagner (C-W) attack \cite{carlini2017towards} has been used to attack the modulation classification task}.
\end{itemize}

The rest of the paper is organized as follows. In the next section, we will provide a brief review of the related research that focuses on ML-based modulation classification and adversarial attacks on modulation classification. Section III describes the methodology where we have discussed the assumed threat model, ML-models used for modulation classification, and the utilized adversarial attack for crafting adversarial examples. Section IV provides the performance evaluation of the adversarial attack on the modulation classification. Section V concludes the study and provides future directions.

\section{Related Work}
\label{rw} 
\subsection{Modulation classification using ML schemes}
The recent success of ML in computer vision and cyber-physical systems has inspired a surge in the utilization of ML schemes in wireless and data networks. It is conceived that ML will be the backbone of future cognitive self-driving networks. Modulation classification is an important problem in dynamic spectrum allocation of cognitive self-driving networks. There are few ML-based modulation classification schemes available in the literature. Wong et al. \cite{wong2004automatic} used a combination of genetic algorithm and multi-layer perceptron for digital modulation recognition. Aslam et al. \cite{aslam2012automatic} used genetic programming with K-nearest neighbor (KNN) for modulation classification. Although genetic algorithms provides a good heuristic-based solution but these algorithms do not scale efficiently with the increase of the sample population.  

Muller et al. \cite{muller2011front} employed a combination of discriminative learning and support vector machines (SVM) for modulation classification. Mendis et al. \cite{mendis2016deep} utilized deep belief networks (DBN) for modulation classification, although DBN has produced very impressive results but they are known to be very difficult to train and scale. O'Shea at al. \cite{o2016convolutional} used convolutional neural network (CNN), VGG, and ResNet for performing modulation classification schemes where they have compared the deep ML-based modulation classification with the conventional modulation schemes under different configuration and noise levels and showed that ML-based schemes performed better even in low signal to noise ratio (SNR). Using ML schemes have produced very good results but they are vulnerable to adversarial examples crafted by the adversary to fool the ML-based classifier to perform incorrect classification. 

\subsection{Adversarial attacks on ML-based modulation classification}
There has not been much work available on exploring the threat of adversarial ML examples on modulation classification. Sadeghi et al. \cite{sadeghi2018adversarial} used a variant of fast gradient sign method (FGSM) \cite{goodfellow2014explaining} to perform an adversarial ML attack on CNN-based modulation classification and successfully showed a considerable drop in classification accuracy. FGSM is a technique for crafting adversarial example where one step gradient update is performed in the direction of the sign associated with the gradient at each feature in the test example. The FGSM perturbation ($\eta$) is given as: 
\begin{equation}\label{eq2}
\eta =\epsilon \textit{sign}(\nabla_x j_\theta(x,l))
\end{equation}

Flowers et al. \cite{flowers2019evaluating} provided an evaluation framework for testing modulation classifiers against adversarial ML attacks, they have tested the modulation classifier against FGSM and Gaussian random noise base adversarial attacks and showed that FGSM causes more destruction than the random Gaussian noise. Similarly, Kokalj et al. \cite{kokalj2019adversarial} used the FGSM attack to demonstrate the vulnerability of modulation classification against adversarial examples. Bair et al. \cite{bair2019limitations} highlighted the limitations of the targeted adversarial attack on modulation classification in white-box settings. 

Most of the results of the adversarial attacks reported on modulation classification have used FGSM attack without considering that FGSM was not designed to generate the optimal amount of adversarial perturbation, it was only designed with an absolute motivation of generating adversarial perturbations quickly rather than optimally \cite{carlini2017towards}. In this paper, we have performed C-W attack \cite{carlini2017towards} on modulation classification to compute the optimal adversarial perturbation.  

\section{Methodology}

In this section, we describe our procedure for performing an adversarial attack on modulation classification. To the best of our knowledge, there is no standardized ML-based solution for modulation classification in the cognitive self-driving networks available yet in the literature, so for completeness we have used both conventional and deep ML schemes for modulation classification. Before delving deep into the details of the ML-models used for modulation classification and adversarial attack on it, we describe the threat model and a few related assumptions. 

\subsection{Threat model} \label{tm}

This subsection describes the major assumptions considered for performing adversarial attack on modulation classifier.

\subsubsection{Adversary Knowledge}
We have assumed a white-box settings for performing an adversarial attack on DNN based modulation classification, which means adversary has the complete knowledge about the model architecture, related hyperparameters, and the test data. This assumption is fairly standard in the adversarial ML domain. We have transferred the adversarial examples for DNN-based modulation classifier to other conventional ML-based modulation classifiers. In this paper, we have only assumed test time adversarial attacks. Poisoning attacks are left for future considerations. 

\subsubsection{Adversarial Goals}
Our goal in this experiment is to compromise the integrity of the modulation classifier through adversarial examples and the success of the adversarial attack in this paper will be measured by the comparison of the accuracy before and after the adversarial attack.

\subsection{Modulation Classification Models}

We have used DNN, KNN, SVM, Na\"ive Bayes (NB), linear discriminant analysis (LDA), Decision Tree (DT), random forest (RF), and ensemble methods for modulation classification. To the best of our knowledge, this is the first paper that uses almost all the famous ML schemes for modulation classification and then performs adversarial ML attack on these schemes to highlight that conventional ML, deep ML, and ensemble methods do not provide robustness against small carefully-crafted perturbations.

\textit{For the DNN classifier}, we have used four dense hidden layers network with rectified linear units as a nonlinear activation in hidden layers and softmax for calculating the classification probabilities of each class. Stochastic gradient descent (SGD) has been used as an optimizer and categorical cross-entropy as the associated loss function for training the DNN based modulation classifier. \textit{For the KNN classifier}, we have used 15 neighbors as an optimal number of neighbors for performing the classification. We have used radial basis function (RBF) kernel for performing the SVM based modulation classification. \textit{For the NB classifier}, we have assumed Gaussian distribution as the underlying modulation data distribution. \textit{For the LDA-based classification}, we have used singular value decomposition (SVD) solver as an SVD solver can better handle a large number of modulation data features. \textit{For the DT classifier}, we have used maximum unfolding depth of 12 for achieving good classification result. \textit{For the RF classifier}, we have a maximum of 10 trees forest for estimating the classification results. \textit{For ensemble methods for modulation classification}, we have employed AdaBoost and Gradient-Boosting algorithms. The obtained classification results are provided in section \ref{ee}.

\subsection{Adversarial attack}
We have performed C-W \cite{carlini2017towards} attack on ML-based modulation classifiers to demonstrate the lack of robustness of the ML-based modulation classification scheme in cognitive self-driving networks. Carlini et al. \cite{carlini2017towards} proposed three very powerful adversarial ML perturbation crafting techniques by using three distance matrices $(L_0, L_2,$ and $L_\infty)$ and these attacks have successfully evaded the \textit{defensive distillation} method \cite{papernot2016distillation} (a popular early scheme for defending against adversarial examples).

We have used $L_2$-based C-W attack for crafting adversarial examples. Instead of formulating the adversarial ML problem as in equation \ref{eq1} (which is highly nonlinear formulation that is difficult to optimize), an alternative formulation (provided in equation \ref{eq3} where $g(x{^*})$ is the new objective function such that $g(x{^*})\leq 0$ iff $g(x^*) = t$ here t can be any label but the true label) is used by the the C-W attack that can be solved by gradient descent. The best performing objective function $g(.)$ used for crafting adversarial examples for modulation classification is provided in equation \ref{eq4}, where $\mathcal{Z}$ denotes the softmax function.   

\begin{equation} \label{eq3}
\begin{split}
    \underset{\eta}{\text{minimize}} \quad \|\eta\|{_\mathcal{P}} + c . g(x{^*}) \\
    \text{such that} \quad (x{^*}) \in [0, 1]{^n}
\end{split}
\end{equation}

\begin{equation} \label{eq4}
    g(x{^*}) = \text{max} \{0, \underset{i \neq t}{\text{max}} \mathcal{Z}(x{^*}){_i} - \mathcal{Z}(x{^*}){_t}\}
\end{equation}

We have only opted to use an $L{_2}$-based adversarial attack because we want to keep the perturbation $\eta$ to a minimum while minimizing the squared error between adversarial modulation example and the original modulation example.

Many defenses against adversarial examples have been proposed in literature but this powerful attack has beaten all of them \cite{carlini2017adversarial} and to the best of our knowledge there does not exist any defense that ensures robustness against ${L_2}$-based C-W adversarial attack.  

In our experiments, we wish to achieve the following objectives:  
\begin{itemize}
\item \textbf{Objective 1:} \textit{Do the ML schemes used for modulation classification in the literature provide necessary robustness against adversarial perturbations?}
\item \textbf{Objective 2:} \textit{We want to experimentally verify that the adversarial examples breaching one ML schemes will breach other ML models with high probability even if the deployed ML model is unknown.}
\end{itemize}
Before explaining how we have met these objectives through our experiment in the next section, we provide a detailed description of the dataset used for performing the experiments.   

\subsection{Dataset} \label{data}
We have used highly cited GNU radio ML RML2016.10a dataset \cite{o2016radio} for our experimentation, the reason for selecting this dataset is its public availability and utilization in the literature. Dataset consists of 220000 input examples, where each example is associated with a modulation scheme at a specific SNR. Dataset has 11 modulation schemes namely: AM-DSB, AM-SSB, WBFM, PAM4, BPSK, QPSK, 8PSK, QAM16, QAM64, CPFSK, and GFSK. Out of these 11 modulation schemes, 8 are digital modulations and 3 are analog modulation schemes.  

For this experiment, we have used eight digital modulation schemes and excluded three analog modulation schemes. The excluded schemes are AM-DSB, AM-SSB, and WBFM. The total number of examples used in these experiments is 160000. Each example is a 256 size vector with 128 in-phase and 128 quadrature-phase components. This dataset was generated for 20 different SNR levels from -20 dB to 18 dB. More details of the dataset preparations can be found in \cite{o2016radio}.

\begin{figure*}[h]
\centering     
\centerline{\includegraphics[width=0.75\textwidth]{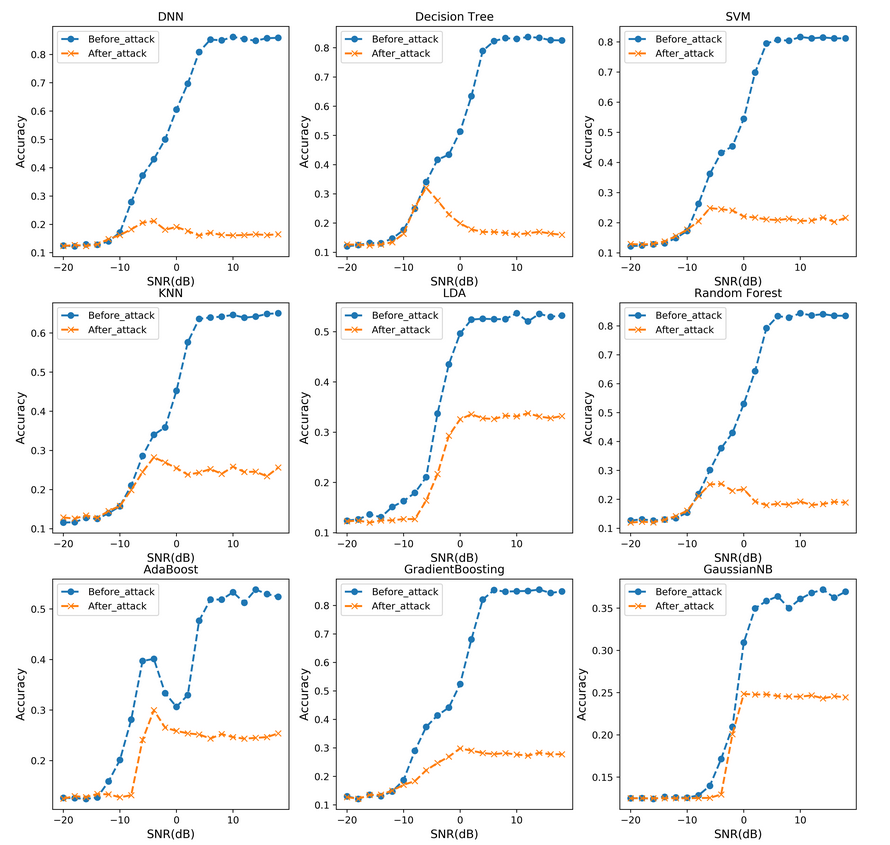}}
\caption{The accuracy of ML models used for modulation classification before and after adversarial ML attack is provided in the figure. A clear drop in the accuracy with improving SNR after the adversarial attack clearly indicates the lack of deterrence against small carefully crafted adversarial perturbations.}
\end{figure*}

\section{Performance Evaluation}
\label{ee}
In this section, we have provided a detailed evaluation of the ML-based modulation classifiers against adversarial perturbations. 

\subsection{Performance Impact}
We have evaluated the ML-based modulation classification before and after the adversarial attack. We have used the accuracy as the performance metric the decay in the modulation classification describes the adversarial attack success. Figure 1 provides a detailed comparison of accuracy and SNR before and after the adversarial attack. The clear drop in the accuracy of the classifiers with increasing SNR fulfills the first objective of this experiment where we set out to show that the ML-schemes proposed in the literature for modulation classification does not provide the necessary robustness against adversarial examples.

\subsection{Transferability of Adversarial Examples}
Here we want to note that adversarial examples were only crafted for DNN-based modulation classifier under white-box assumptions. The adversarial examples compromising the integrity of the DNN classifiers were transferred to the rest of ML classifiers under black-box assumptions and it turns out that modulation classifiers based on conventional ML techniques are also equally vulnerable to the adversarial examples which fulfill the second objective we wanted to achieve through this experiment.  

\section{Conclusions}
In this paper, we have highlighted the lack of robustness of ML-models utilized in modulation classification by successfully evading 9 different ML-based modulation classifiers. We have also successfully shown that transferability of the adversarial examples from one model to another model for performing the adversarial attack. This work has also provided a glimpse of the security and robustness issues associated with the utilization of ML models in cognitive self-organizing networks. Designing new defenses against adversarial attacks on cognitive self-driving networks are left as a future direction.   
   
\def\bibfont{\small}
\bibliographystyle{IEEEtranS}
\bibliography{IEEEabrv,reference}

\end{document}